\documentclass[a4paper,12pt]{article}
\usepackage[utf8]{inputenc}
\usepackage[T1,T2A]{fontenc}
\usepackage{amsmath,amsfonts,amssymb}
\usepackage{cite}

\newcommand{\lb}[0]{\left(}
\newcommand{\rb}[0]{\right)}
\newcommand{\lsb}[0]{\left[}
\newcommand{\rsb}[0]{\right]}

\begin{document}

\renewcommand*{\thefootnote}{\fnsymbol{footnote}}

\begin{center}
{\large\bf Confinement potential in Soft Wall holographic approach to QCD
%\\[0.25cm]
%\foreignlanguage{russian}{Потенциал конфайнмента в голографическом подходе с <<мягкой стенкой>> к КХД}
}
\end{center}
%\bigskip
\vspace{-0.1cm}
\begin{center}
{Sergey Afonin\footnote{E-mail: \texttt{s.afonin@spbu.ru}},
Timofey Solomko%\footnote{e-mail: \texttt{tsolomko@gmail.com}}
}
\end{center}

\renewcommand*{\thefootnote}{\arabic{footnote}}
\setcounter{footnote}{0}

\begin{center}
{\small Saint Petersburg State University, 7/9 Universitetskaya nab.,
St.Petersburg, 199034, Russia}
\end{center}

\bigskip

{\footnotesize
\noindent We present the results for the confinement potential of the Cornell
type within the framework of the generalized Soft Wall holographic model
(with quadratic dilaton background in the metric) which contains an additional
parameter responsible for the value of intercept of the linear Regge spectrum.
Next, we note that the Cornell potential arises also in some Soft Wall models which do not
lead to Regge-like spectrum. As an example, we demonstrate this property in a model with
linear dilaton background in the metric. This example is at odds with intuition based on the hadron string picture
that the linearly rising potential and Regge-like spectrum are directly related.
%\\[0.25cm]
%\foreignlanguage{russian}{Представлены результаты для потенциала конфайнмента
%корнельского типа в рамках обобщённой голографической модели с мягкой стенкой (с квадратичным дилатонным фоном в метрике), которая
%включает параметр, отвечающий за величину <<интерсепта>> линейного реджевского спектра.
%Далее отмечено, что корнельский потенциал возникает также в некоторых моделях с мягкой стенкой,
%не ведущих к реджевскому спектру. В качестве примера рассмотрена модель с линейным дилатонным фоном в метрике.
%Этот пример противоречит интуиции, основанной на картине адронной струны,
%что линейно растущий потенциал и Редже-подобный спектр напрямую связаны.
%}\\[0.25cm]
%PACS: 12.38.-t
}

\section*{Introduction}

The heavy-quark potential represents one of the key observables that is related to confinement.
The lattice simulations of its form (see, e.g., the review~\cite{bali}) yield
\begin{equation}\label{cornell}
  V(r)=-\frac{\kappa}{r}+\sigma r + \text{const},
\end{equation}
which is in a remarkable agreement with the Cornell potential. %~\cite{Eichten:1978tg}.
Compliance with this result is an important test of the viability of various
phenomenological approaches to strong interactions.

The so-called Soft Wall (SW) holographic model is one of the models that successfully passes
this check~\cite{son2,andreev}. The heavy-quark potential was first calculated within the
SW holographic model by Andreev and Zakharov in~\cite{Andreev:2006ct} and their result
agreed with~\eqref{cornell}. The analysis in~\cite{Andreev:2006ct}, however, was
performed only for a particular case of the vector SW model with a fixed intercept of
string like mass spectrum. Recently, this method was successfully extended to the cases of
the generalized~\cite{Afonin:2021cwo} vector and scalar SW models
in~\cite{Afonin:2021zdu} and~\cite{Afonin:2022aqt}. However, in situations where the Regge
mass spectrum is not a strict requirement one may consider other types of the dilaton
background in the holographic model that may still provide correct behavior of the
potential.

In this report, we first briefly describe the idea of holographic derivation of the potential
and review the existing results provided by the generalized SW model. After that we
consider one of the possible alternatives, specifically, the so-called ``linear dilaton''
background. We present the results for the potential in this case and compare them with
the standard case of SW model with ``quadratic dilaton''.

\section*{The holographic setup and Wilson loop}

The standard SW holographic model is defined by the 5D action~\cite{son2}
\begin{equation}
\label{sw}
  S=\int d^4\!x\,dz\sqrt{g}\,e^{-cz^2}\mathcal{L},
\end{equation}
where $g=|\text{det}g_{MN}|$, $\mathcal{L}$ is a Lagrangian density of some
free fields in AdS\(_5\) space which, by assumption, are
dual on the AdS\(_5\) boundary to some QCD operators.
The metric is given by the Poincar\'{e} patch of the AdS$_5$ space,
\begin{equation}
\label{2}
g_{MN}dx^Mdx^N=\frac{R^2}{z^2}(\eta_{\mu\nu}dx^{\mu}dx^{\nu}-dz^2),\qquad z>0.
\end{equation}
Here $\eta_{\mu\nu}=\text{diag}(1,-1,-1,-1)$, $R$ denotes the radius of AdS$_5$ space,
and $z$ is the holographic coordinate. The 4D mass spectrum of this model can be found
from the equation of motion accepting the 4D plane-wave ansatz.
For instance, in the case
of free massless vector fields, one considers the particle-like ansatz
$V_\mu(x,z)=e^{ipx}v(z)\epsilon_\mu$ with the on-shell, $p^2=m^2$, and transversality,
$p^\mu\epsilon_\mu=0$, conditions~\cite{son2}. The resulting spectrum takes the Regge form,
\begin{equation}
\label{8}
m_n^2=4|c|n,\qquad n=1,2,\dots.
\end{equation}

The generalized SW holographic model describes the linear spectrum with arbitrary intercept regulated by a parameter $b$.
In the vector case, the generalization of spectrum~\eqref{8} is
\begin{equation}
\label{spSW}
m_n^2=4|c|(n+b),\qquad n=1,2,\dots.
\end{equation}
As was shown in~\cite{Afonin:2012jn} and further developed in~\cite{Afonin:2021cwo},
the spectrum~\eqref{spSW} arises in the following generalization of the action of the
vector SW model,
\begin{equation}
\label{gen_sw}
  S=\int d^4xdz\sqrt{g}e^{-cz^2}U^2(b,0,|cz^2|)\mathcal{L},
\end{equation}
where $U$ is the Tricomi hypergeometric function that modifies the dilaton background.

Both of these models can be reformulated in such a way that the dilaton background is
eliminated and instead the AdS$_5$ metric is modified. For the standard SW model such
reformulation was proposed in~\cite{andreev}. It reads
\begin{equation}\label{az_metric}
  g_{MN}=\text{diag}\left\lbrace\frac{R^2}{z^2}h,\dots,\frac{R^2}{z^2}h\right\rbrace,\quad
  h=e^{-2cz^2},
\end{equation}
and most notably it leads to the same equation of motion, i.e., it does not change the discrete spectrum.

The generalized SW model can be also rewritten in a similar way, by observing that if
a 5D holographic model in its action contains a $z$-dependent background described by a
function $B(z)$ then the action can be rewritten in the form
\begin{equation}\label{ac2}
  S=\int d^4x dz\sqrt{\tilde{g}}\,\tilde{\mathcal{L}},
\end{equation}
with the modified metric~\cite{Afonin:2021cwo}
\begin{equation}\label{tr}
  \tilde{g}_{MN}=B^2g_{MN}.
\end{equation}
Substituting the background $B(z)$ in the action~\eqref{gen_sw}, we obtain the following
generalization for the function $h$ in the modified metric~\eqref{az_metric},
\begin{equation}\label{gen_h}
  h=e^{-2cz^2}U^4(b,0,|cz^2|).
\end{equation}
This background function serves as a starting point in the derivation of the potential
within the generalized SW holographic model.

The procedure for the derivation of the potential at its core is based on the analysis of
Wilson loop. Within the holographic framework, this analysis was proposed by Maldacena
in~\cite{Maldacena:1998im}. There one considers a Wilson loop \(W\) situated on
the 4D boundary of 5D space with the time coordinate ranging from \(0\) to \(T\) and
the remaining 3D spatial coordinates \(y\) from \(-r/2\) to \(r/2\). The expectation value
of the loop in the limit of \(T\to\infty\) is equal to \(\left\langle W\right\rangle\sim e^{-TE(r)}\),
where \(E(r)\) is the energy of the quark-antiquark pair. On the other hand, this
expectation value can be obtained via \(\left\langle W\right\rangle\sim e^{-S}\),
where \(S\) represents the area of a string world-sheet which produces the loop
\(W\). Combining these two equations the energy of the configuration is computed
simply as \(E=S/T\). The natural choice for the world-sheet area is the Nambu-Goto
action and in what follows \(\alpha'\) denotes the inverse string tension appearing in
this action.

The specific application of this idea to the case of the SW holographic model
was developed in~\cite{Andreev:2006ct} and applied further to the case of the generalized
SW model in~\cite{Afonin:2021zdu} and~\cite{Afonin:2022aqt}. The final results are
presented in the form of the asymptotics of the potential at either large or small
distances. The exhaustive details of the derivation of the results may be found
in~\cite{Afonin:2021zdu} and~\cite{Afonin:2022aqt} and below we only summarize the
final outcome of this procedure.

In the case of the generalized vector SW model at large distances we have a linear
potential
\begin{equation}\label{pot_large_r}
  E\underset{r\to\infty}{=}\frac{R^2}{\alpha'}\sigma_\infty r,\quad
  \sigma_\infty=\frac{e^{2x}U^4(b,0,x)}{2\pi x}c,
\end{equation}
where $x=x(b)$ represents a non-trivial function of $b$ defined by the equation arising
from the requirement that certain integral expressions for the energy and the distance
take only real values (see~\cite{Afonin:2021zdu} for more details). At small distances
one obtains ``linear plus Coulomb'' asymptotics,
\begin{equation}
\label{pot_small_r}
  E\underset{r\to0}{=}\frac{R^2}{\alpha'}\lb-\frac{\kappa_0}{r}+\sigma_0 r\rb,
\end{equation}
\begin{equation}
\label{small_r_not}
  \kappa_0\equiv\frac{(2\pi\rho^2)^{-1}}{\Gamma(1+b)^4},\quad
  \sigma_0\equiv\frac{1+2b\psi(b+1)+4b\gamma}{\Gamma(1+b)^4}c\rho^2,
\end{equation}
where \(\rho\equiv\Gamma(1/4)^2/(2\pi)^{3/2}\).

Similar analysis may be applied to the scalar case of the generalized SW model where the
action takes the form~\cite{Afonin:2021cwo},
\begin{equation}\label{gen_sw_s}
  S_\text{sc}=\int d^4xdz\sqrt{g}e^{-cz^2}U^2(b,-1,|cz^2|)\mathcal{L}_\text{sc},
\end{equation}
where \(\mathcal{L}_\text{sc}\) is a Lagrangian of a scalar 5D field. The resulting
potential asymptotics has the same form of the dependence on \(r\),
i.e.,~\eqref{pot_large_r} and~\eqref{pot_small_r}, but with modified coefficients,
\begin{equation}
  \sigma_\infty=\frac{e^{2x/3}U^{4/3}(b,-1,x)}{2\pi x}c,\quad
  \kappa_0=\frac{(2\pi\rho^2)^{-1}}{\Gamma(2+b)^{4/3}},\quad
  \sigma_0=\frac{1-2b}{\Gamma(2+b)^{4/3}}\frac{c\rho^2}{3}.
\end{equation}
where \(x\) is defined via the scalar counterpart of the equation mentioned above.

\section*{SW model with linear dilaton}

It is well-known that the quadratic structure of the dilaton exponential background is
necessary to obtain the correct Regge behavior of meson mass spectra, \(m_n^2\sim n\). Let us,
however, set aside for now this requirement and consider the case of what we shall call
``linear dilaton''\footnote{The SW-like holographic models with asymptotically linear dilaton were first
proposed in~\cite{Afonin:2009xi} as a holographic way for description of radial meson spectra with a finite
number of states.}, i.e. (note the different power in the exponent),
\begin{equation}\label{act}
  S=\int d^5xe^{cz}\sqrt{g}\mathcal{L}_\text{sc},
\end{equation}
where \(\mathcal{L}_\text{sc}\) corresponds to a Lagrangian of a scalar 5D field. The mass
spectrum of this model
takes the form (parameter \(k\) is defined as a solution to the equation \(k(k-1)=15/4+m_5^2R^2\),
where \(m_5\) is the mass of the scalar 5D field)
\begin{equation}
  m_n^2=\frac{c^2}{4}-\frac{9c^2}{16(n+k-1)^2},\qquad n=0,1,2,\dots,
\end{equation}
which is in a stark contrast to the Regge-like spectra~\eqref{8} and~\eqref{spSW} of SW models with
quadratic dilaton background. In terms of the modified AdS metric, the function \(h(z)\) in
this case is equal to
\begin{equation}
  h(z)=e^{2cz/3}.
\end{equation}
Applying the same holographic Wilson procedure for calculating asymptotics of static potential
we obtained the following potential at small distances,
\begin{equation}
  E\underset{r\to0}{=}\frac{R^2}{\alpha'}
  \lsb-\frac{1}{2\pi\rho^2}\frac{1}{r}+\frac{c\ln2}{3\pi}+
  \frac{c^2\rho}{18\pi}\lb\lb3\pi+4\rb\rho-4\rb r\rsb,
\end{equation}
and at large distances,
\begin{equation}
  E\underset{r\to\infty}{=}\frac{R^2}{\alpha'}\frac{c^2e^2}{18\pi}r.
\end{equation}

Comparing these results with the previous cases we can make a couple of observations.
First, at small distances the usual Coulomb and linear terms are reproduced, but in addition to
that an explicit constant term emerges. Second, at large distances the form of the
potential remains linear. Note that the numerical ratio of the linear term coefficient at large
distances to small distances is \(\sigma_\infty/\sigma_0\approx1.23\) which is almost equal to the result
of standard SW model with quadratic dilaton background (namely \(\sigma_\infty/\sigma_0\approx1.24\)~\cite{Andreev:2006ct}).

\section*{Conclusion}

In this report we briefly reviewed the calculation of static potential arising between static sources
within the framework of generalized SW models for strong interactions. We also considered a particular
case of other type of the dilaton background, namely the SW model with ``linear dilaton''. The spectrum of
the second model is not Regge-like. However, the static potential predicted by this model was shown to
be qualitatively the same, i.e., of the Cornell type. This means that within the bottom-up holographic
approach to strong interactions, the linearly rising potential does not necessarily entail a Regge-like spectrum,
in contrast to the hadron string picture.

It seems that the static potential linearly rising at large distances is a general feature of any SW-like
holographic model satisfying the Sonneschein confinement criterion~\cite{Sonnenschein:2000qm}.
An analysis of this issue is left for the future work.

\section*{Acknowledgements}

This research was funded by the Russian Science Foundation grant number 21-12-00020.

\end{document}